**Title:** Active tuning of plasmon damping via light induced magnetism.

**Author list:**
Oscar Hsu-Cheng Cheng[1]*, Boqin Zhao[1]*, Zachary Brawley[3], Dong Hee Son[1,2], Matthew Sheldon[1,3]
[1]Department of Chemistry, Texas A&M University, College Station, TX, USA.
[2] Center for Nanomedicine, Institute for Basic Science and Graduate Program of Nano Biomedical Engineering, Advanced Science Institute, Yonsei University, Seoul, Republic of Korea.
[3]Department of Material Science and Engineering, Texas A&M University, College Station, TX, USA.
*These authors contributed equally to this work.
e-mail: dhson@chem.tamu.edu; sheldonm@tamu.edu

**Abstract**

Circularly polarized optical excitation of plasmonic nanostructures causes coherent circulating motion of their electrons, which in turn, gives rise to strong optically induced magnetization— a phenomenon known as the inverse Faraday effect (IFE). In this study we report how the IFE also significantly decreases plasmon damping. By modulating the optical polarization state incident on achiral plasmonic nanostructures from linear to circular, we observe reversible increases of reflectance by 78% as well as simultaneous increases of optical field concentration by 35.7% under $10^9$ W/m$^2$ continuous wave (CW) optical excitation. These signatures of decreased plasmon damping were also monitored in the presence of an externally applied magnetic field (0.2 T). The combined interactions allow an estimate of the light-induced magnetization, which corresponds to an effective magnetic field of ~1.3 T during circularly polarized CW excitation ($10^9$ W/m$^2$). We rationalize the observed decreases in plasmon damping in terms of the Lorentz forces acting on the circulating electron trajectories. Our results outline strategies for actively modulating intrinsic losses in the metal, and thereby, the optical mode quality and field concentration via opto-magnetic effects encoded in the polarization state of incident light.

**Introduction**

The reversible modulation of plasmonic resonances in metal nanostructures using external stimuli – i.e., "active plasmonics"— is currently of great interest for potential applications in sensing, optoelectronic devices, and light-based information processing[1]. Commonly explored strategies modulate plasmon resonance frequencies by altering the surrounding dielectric environment of nanostructures, for example, with thermoresponsive materials[2-4]. Modification of plasmonic modes based on distance-dependent optical coupling between nanostructures in compliant media under stress and strain has also been demonstrated[5-7]. Alternatively, the optical properties of the metal comprising a nanostructure can be reversibly modulated. Within the Drude model, the complex dielectric function of a metal at angular frequency, $\omega$, is well described in terms of the electrical carrier density, $n$, and the damping constant, $\gamma$, for the carrier oscillations:

$$\varepsilon(\omega) = 1 - \frac{\omega_p^2}{\omega^2 + i\omega\gamma} \quad (1)$$

where $\omega_p = \sqrt{\frac{ne^2}{\varepsilon_o m_e}}$ is the bulk plasma frequency, also depending on the electron charge, $e$, effective mass, $m_e$, and vacuum permittivity, $\varepsilon_o$ [8]. Researchers have shown that capacitive surface charging of metals when they are integrated into electrochemical cells results in reversible shifts of their plasmon resonance frequency through the dependence on $n$ [9-11]. In the time domain, pulsed laser excitation can similarly perturb electronic carrier populations giving rise to transient modulation of plasmonic behavior [12-17].

In comparison, the possibility of actively tuning plasmon damping in the steady state, and the opportunities for manipulating plasmonic behavior, has been studied much less [18-21]. In equation (1), $\gamma$ reflects several different microscopic processes connected with the conductivity and mean free path of electrical carriers in the metal such as electron-electron scattering, electron-phonon scattering, surface scattering [22], and chemical interface damping with surface adsorbates [23], in addition to other loss pathways such as radiation damping [24] and Landau damping [25,26]. Usually, $\gamma$ is considered to be an intrinsic property that is determined by the chemical identity of the metal [27,28], surface chemistry and morphology, such as the crystal grain size [29-32], or the modal behavior at a particular frequency [19,33], e.g. near field localization versus far field out-coupling. However, changes in plasmon damping have a profound impact on the ability of a nanostructure to concentrate optical power in a particular mode, known as the quality factor or "Q" factor. Decreasing $\gamma$ lowers the imaginary part of the permittivity, decreasing ohmic losses from carrier motion at the optical frequency and increasing overall optical scattering or reflectance. Lower damping also provides greater field enhancement at sub-wavelength "hot spots", improving efficiency for localized sensing, photochemistry, or heating via photothermalization.

In a series of recent studies, the Vuong laboratory reported anomalously large magneto-optical (MO) responses from colloidal Au nanoparticles under small magnetic fields (~1 mT) and low intensity circularly polarized (CP) excitation (<1 W/cm$^2$) [34,35]. The large MO response was attributed to the interaction between external magnetic fields and circulating drift currents in the metal that were resonantly excited during CP excitation (Fig. 1a). The generation of drift currents from coherent charge density waves that circulate in metals during CP optical cycles has been studied extensively theoretically [36-38], and is understood to contribute to optically induced static (DC) magnetism. This behavior is also known as the Inverse Faraday Effect (IFE) [36-41]. Our recent experimental study of 100 nm Au nanoparticle colloids showed ultrafast modulation of effective magnetic fields up to 0.038 T under a peak intensity of 9.1×10$^{13}$ W/m$^2$, confirming MO activity and optically-induced magnetism many orders of magnitude greater than for bulk Au [42]. Notably, Vuong et al. also reported an apparent increase of the volume averaged electrical conductivity when an external magnetic field was aligned with the light-induced magnetic field. Considering several recent theoretical and experimental magnetic circular dichroism (MCD) studies [43-45], these results can be interpreted as a decrease in plasmonic damping when the microscopic electron motion in a plasmon resonance contributes to DC magnetization.

In this study we show that plasmon damping is indeed strongly modified by magnetic interactions, whether magnetism is induced externally using an applied magnetic field or created internally with light via the IFE. We observe that the normal incidence backscattering, i.e. the reflectance, of arrays of non-chiral Au nanostructures is increased by 78% when controlling the ellipticity of incident light from linearly polarized (LP) to CP during $10^9$ W/m$^2$ continuous wave (CW) excitation. Further, we query the optical field concentration at hot spots in the metal by taking advantage of recently refined electronic Raman thermometry techniques[46-53]. Localized photothermal heating induces temperatures ~23 K greater during CP excitation than during LP excitation under an excitation intensity of ~$10^9$ W/m$^2$, suggesting active modulation of optical field enhancement by 35.7% at hot spots. The local heating is further modulated by the presence of an externally applied magnetic field (~0.2 T), supporting the underlying magnetic origin of these phenomena and allowing estimation of the light-induced effective magnetic fields at hot spots (~1.3 T). Taken together, our results indicate reversible modulation of the plasmon damping that can exceed ~50%, solely by controlling the ellipticity of the incident radiation.

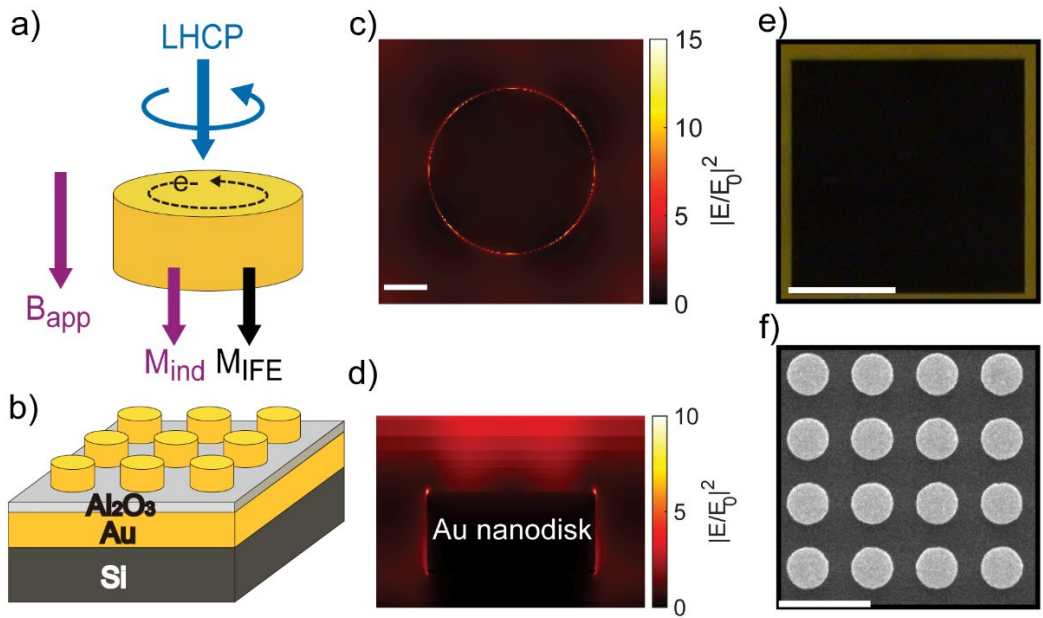

**Fig. 1 Sample overview.** (a) Relationship between incident light helicity, induced electronic motion, induced magnetization from the IFE ($M_{IFE}$), and induced magnetization ($M_{ind}$) from an external static magnetic field ($B_{app}$) in a plasmonic nanostructure. When incident light has left-handed circular polarization (LHCP), $M_{IFE}$ and $M_{ind}$ are parallel. (b) Schematic of the Au nanodisk array sample (not to scale). (c) Top view and (d) side view of the local field enhancement during 532 nm excitation. Width and height not to scale in (d). (e) Optical image of the array on an Au film (f) SEM image. Scale bars: (c) 100 nm (e) 40 µm and (f) 1 µm.

**Result and Discussion**

Samples consisting of 100 µm x 100 µm arrays of 400 nm diameter by 100 nm height disk-shaped gold nanostructures in a square lattice pattern (700 nm pitch) were deposited on 38 nm thick $Al_2O_3$ layer on top of 100 nm thick gold films using electron-beam lithography (Fig. 1b, e, f, see Methods section). The nanodisk shape supports circulating electronic currents during CP excitation (Fig. 1a). Periodic arrays provide high absorptivity across the visible spectrum (Fig. 1e, Fig. 2a), aiding photothermal heating for the Raman thermometry studies detailed below. The overall sample geometry is achiral, highly symmetric, and exhibits no polarization or ellipticity dependence for absorption or scattering (neglecting non-linear effects), as confirmed by full wave optical simulations (FDTD method, see Supporting Information).

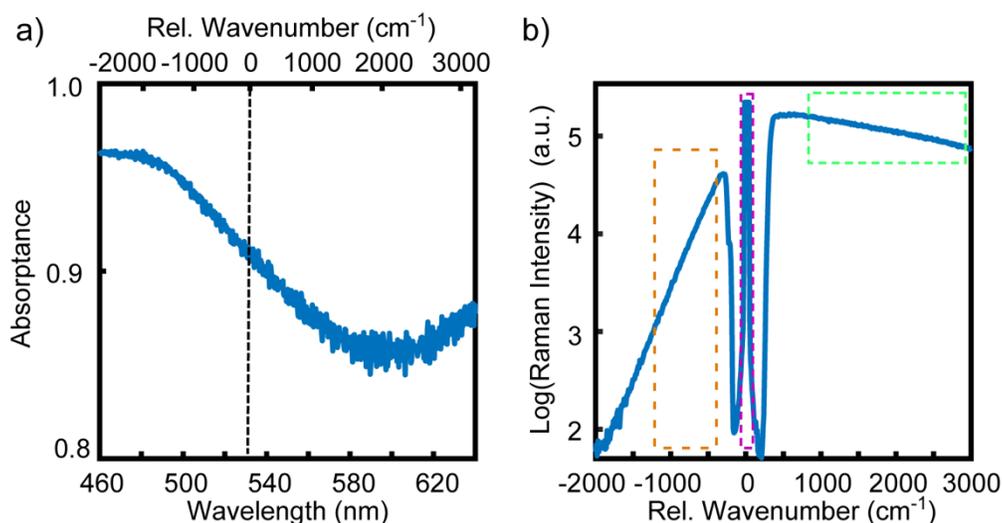

**Fig. 2 Spectroscopy of gold nanodisk arrays.** (a) Absorption spectrum. The dashed line indicates 532 nm. (b) Electronic Raman spectrum during 532 nm CW excitation. Different spectral regions provide information analyzed in this study. Orange box: anti-Stokes signal used for temperature fitting. Purple box: 532 nm backscattering (Rayleigh line, filtered here) used to quantify ellipticity-dependent reflection. Green box: the broad energy distribution of the Stokes region is fit for the plasmon damping, $\gamma$.

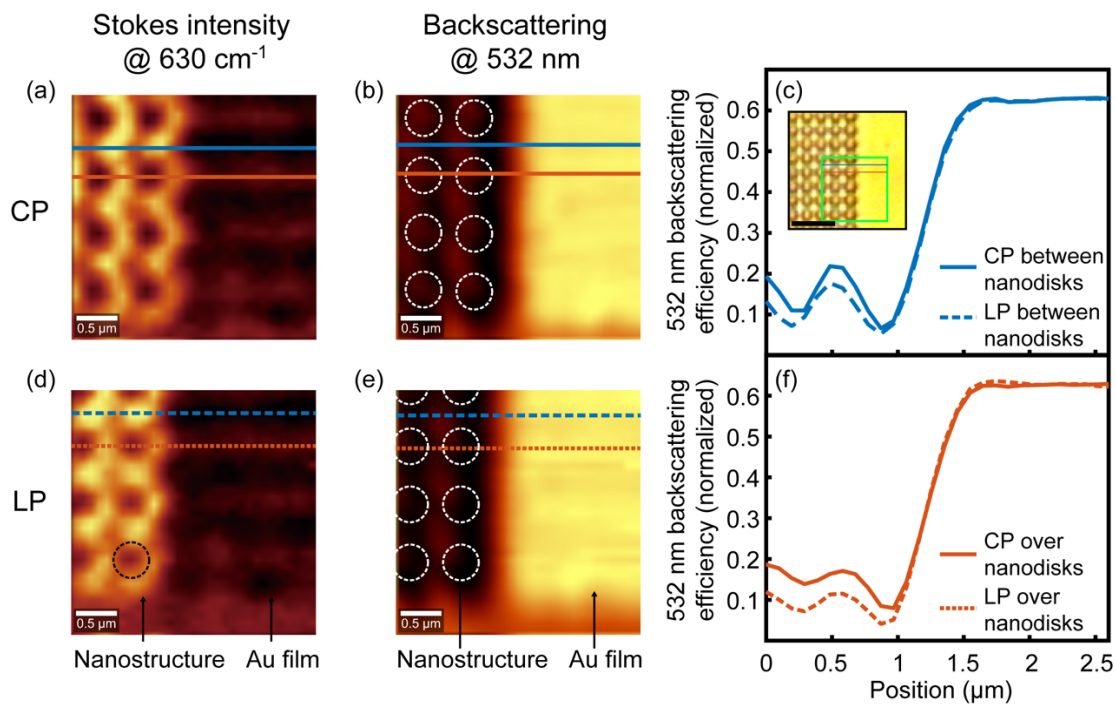

**Fig. 3 Confocal spectral mapping under different helicities.** (a, d) Confocal Raman and (b, e) backscattering intensity map of the gold nanodisk array under 532 nm CW excitation with (a, b) CP or (d, e) LP. (c, f) Line scans of the backscattering efficiency along the region between (blue) or over (red) nanodisks with LP (dashed trace) or CP (solid trace) excitation. Inset: optical image of the sample array. The green box indicates the region of the confocal map. Scale bar: 4 μm.

Samples were mounted onto a piezo-driven microscope stage, and confocal maps of the electronic Raman (eR) spectrum (Fig. 3a, d) or the backscattering intensity at 532 nm (Fig. 3b, e) were collected as a function of position over the edge of an array during linearly polarized (LP) or circularly polarized (CP) excitation. A representative eR spectrum is shown in Fig. 2b. In comparison with typical Raman signals that result from inelastic scattering with vibrational modes in a sample, the eR signal is due to inelastic interactions with the electron gas at the metal surface. The broad eR signal therefore provides information about the energetic distribution of electrons, such as their temperature. As clearly seen by comparison of Fig. 3a, d, and Fig. 1c, d, the eR signal is strongest at "hot spots" at the edge of nanodisks where field enhancement is greatest. In contrast, the sample backscattering, or reflectance, (Fig. 3b, e) is lower over the array due to pronounced plasmonic absorption, ultimately giving rise to localized photothermal heating.

The backscattering efficiency of the nanodisks is strongly modulated based on the polarization state (ellipticity) of the incident light. As displayed in Fig. 3c, f, we compared the backscattered light intensity from different locations over the nanodisk array and the adjacent Au film during LP (dashed trace) or CP (solid trace) excitation. The signal intensity was converted to backscattering efficiency by referencing the Au film region to a smooth Ag mirror, in order to rule out any polarization-dependent instrument response.

(see Methods section for details). In all locations over the array the backscattering efficiency is generally larger for CP versus LP, with a maximum relative increase of backscattering up to 78% at the edge of individual nanodisks during CP excitation.

We rule out the possibility that this trend is due to inherent differences in the ellipticity-dependent scattering efficiency based on sample geometry, because the sample is not chiral. Neglecting optically induced magnetization or other non-linear effects, the total absorption and scattering of the nanostructure array is expected to show no dependence on beam ellipticity. Indeed, we have performed linear, full wave optical simulations (FDTD method, see Supporting information) that confirm no difference in the absorption or scattering efficiency based on LP or CP excitation. We hypothesize that the large difference in backscattering efficiency observed experimentally results from ellipticity-dependent modulation of the plasmon damping. This interpretation is qualitatively supported by additional optical simulations (see Supporting information) that show a comparable increase of backscattering, depending on sample location, when damping is decreased by 50% or greater.

Decreased plasmon damping also results in more concentrated optical fields at hot spots, which can lead to more pronounced local photothermal heating. In terms of equation (1), as the damping constant $\gamma$ decreases the imaginary part of the dielectric function $Im(\varepsilon(\omega))$ also decreases, giving rise to a larger Q-factor and greater local field enhancement[54]. The local power density for heating, $q(x,y,z)$, depends on field enhancement as[55,56]:

$$q(x,y,z) = \frac{1}{2} Im[\varepsilon(x,y,z)]\varepsilon_0 |E(x,y,z)|^2 \quad (2)$$

where $E(x,y,z)$ is the local electric field. Thus, lower plasmon damping provides a net increase of heating power and correspondingly larger temperatures at locations with strong field enhancement. Intuitively, lower damping increases the cross section that funnels light energy into a plasmonic hot spot. See simulations of this effect for the nanodisk array geometry in Supporting information.

Raman signal intensity also depends on local field enhancement, scaling as $|E|^4$ [57]. Therefore Raman-based thermometry techniques primarily probe the nanostructure temperature at hot spots. We measured the sample temperature at hot spots by adapting an anti-Stokes (aS) Raman thermometry method developed by Xie et al[46]. Experimentally, it has been shown that the spectral intensity of the aS eR signal, $S(\Delta\omega)$, is thermally activated according to a Bose-Einstein distribution. The aS spectrum collected from a sample at an unknown temperature, $T_l$, can be normalized by a spectrum collected at a known temperature, $T_0$, according to

$$\frac{S(\Delta\omega)_{T_l}}{S(\Delta\omega)_{T_0}} = \frac{exp\left(\frac{-hc\Delta\omega}{k_B T_0}\right)-1}{exp\left(\frac{-hc\Delta\omega}{k_B T_l}\right)-1} \quad (3)$$

where $h$ is Plank constant, $c$ is the speed of light, $\Delta\omega$ is the wavenumber (negative for anti-Stokes), and $k_B$ is Boltzmann constant. Spectral features that do not change with thermal activation, such as the frequency-dependent signal enhancement factor, cancel out, so that $T_l$ is the only unknown fitting parameter.

With fixed linear polarization, we measured the aS spectra of samples as a function of excitation power to induce variable amounts of photothermal heating, and we fit for $T_l$. The spectra were normalized by a spectrum collected at the lowest possible power that preserved good signal-to-noise (I = 7.2 × 10$^6$ W/m$^2$), with the goal of inducing minimal heating above room temperature, i.e. $T_0 \approx 298$ K. We observe that $T_l$ increases linearly with LP excitation intensity (Fig. 4, blue line), in good agreement with many other reports of gold and copper nanostructures[46,50,58]. The linear fit to the temperature trend shows a y-intercept near room temperature, at the limit of zero incident power, further confirming the accuracy of the thermometry technique. The fitted slope of the trend (4.6×10$^{-8}$ K·m²/W) describes the "heating efficiency" of the sample under LP excitation.

We determined changes in the sample temperature, and hence modification of the plasmon damping, by measuring the eR response while varying the ellipticity of the excitation beam. However, given the complex spectral dependence of the eR signal, similar procedures as those described for the backscattering study could not be used to correct for the ellipticity-dependent instrument response. Instead, we devised a dual beam configuration (see Methods section for details). In summary, two separate CW 532 nm laser beams were coincident on the sample. A low power "Beam 1" was maintained with linear polarization. A second, higher power "Beam 2" was used to induce variable magnetization and damping in the sample by controlling excitation ellipticity. The eR spectrum resultant from Beam 1 was isolated and fit for $T_l$ by collecting a spectrum with both Beam 1 and Beam 2 incident, and then subtracting the spectrum collected with only Beam 2 incident. This procedure allowed us to probe the sample temperature using a beam that had non-changing incident power and linear polarization (Beam 1) while the sample was excited with variable ellipticity and power. The accuracy of this dual-beam method is confirmed in Fig. 4 (red trace). The fitted $T_l$ are reported as a function of the total incident power of Beam 1 and Beam 2, with the power of Beam 1 held constant and both beams maintained in linear polarization. A similar y-intercept and heating efficiency is observed using either thermometry technique.

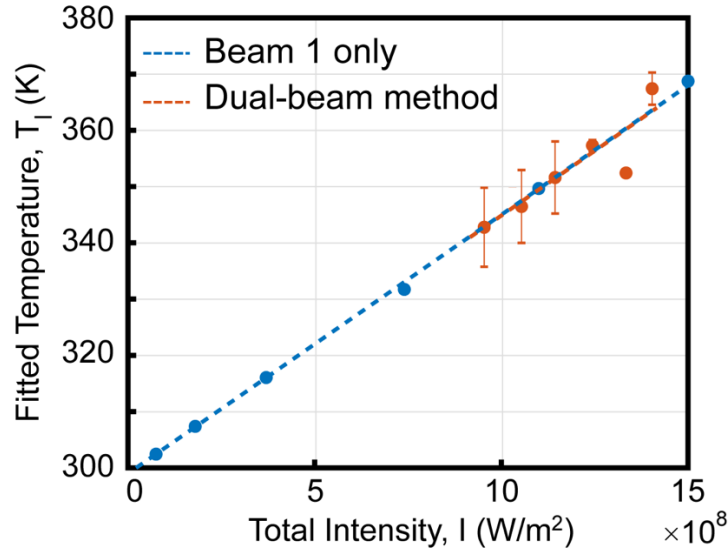

**Fig. 4 Photothermal heating with a single- or dual-beam geometry.** (a) Fitted nanostructure temperature, $T_l$ as function of excitation intensity, $I$, when only Beam 1 (blue) or when both Beam 1 and Beam 2 (red) were incident. Blue linear fit: $T_l (K) = 4.6 \times 10^{-8} I + 299$. For the dual-beam study the intensity of Beam 1 was kept at 4.8×10$^8$ (W/m²), and the intensity of Beam 2 varied between 6.6 – 9.5×10$^8$ (W/m²). Red linear fit: $T_l (K) = 4.6 \times 10^{-8} I + 299$.

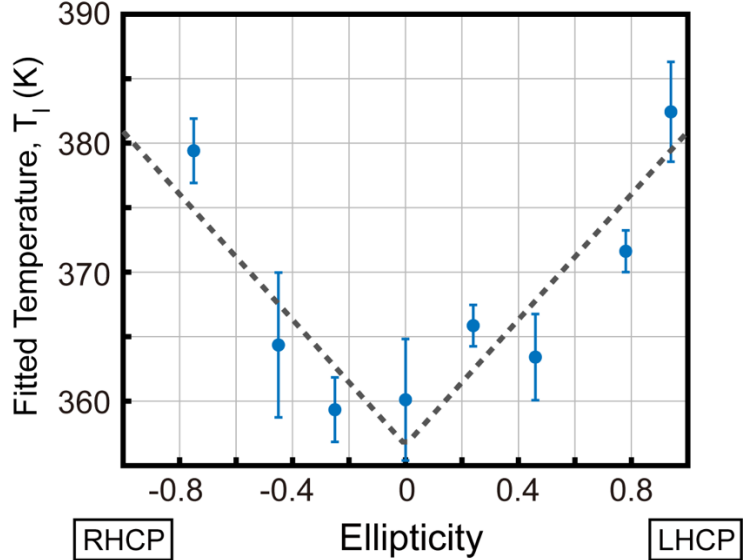

**Fig. 5. Ellipticity dependent photothermal heating.** Fitted nanostructure temperature as a function of Beam 2 ellipticity. The intensity of Beam 1 was kept at 4.8×10⁸ W/m². The intensity of Beam 2 was kept at 9.5×10⁸ W/m². The dashed line is a guide for the eye. See Fig. S8 for a detailed explanation of ellipticity and ellipticity angle.

We then examined the dependence on the ellipticity of Beam 2 while the power of both beams was held constant (Fig. 5). For both right-handed or left-handed circular polarization (RHCP or LHCP), $T_l$ increases with increasing ellipticity. For the same magnitude ellipticity but opposite helicity, the increase of $T_l$ is similar. This trend indicates enhanced field concentration at hot spots when the ellipticity-dependent magnetization in the sample is increased. The maximum increase of temperature observed for CP compared with LP is ~23 K. Based on the heating efficiency determined under LP excitation (Fig. 4a, red line), this same temperature increase would be expected if the sample received 5 × 10⁸ W/m² more incident power. Since the total excitation power was kept constant at 1.4 × 10⁹ W/m², we conclude that the switch from LP to CP excitation equivalently increases the heating power at hot spots by 35.7%. We emphasize that field enhancement is not expected to depend on excitation ellipticity, neglecting non-linear effects, because the sample is not chiral.

We also studied sample behavior under an externally applied magnetic field. For this experiment the ellipticity of Beam 2 was kept at either 0 (LP), +0.67 (LHCP), or -0.67 (RHCP). Note that the ellipticity for RHCP was limited to this range based on our experimental geometry (see Methods section). An external magnetic field $B_{app}$ = 0.2 T was applied parallel to the direction of light propagation (Fig. 1a), and $T_l$ was measured as a function of total incident power. As summarized in Fig. 6, the fitted $T_l$ are larger under LHCP excitation compared to RHCP, while both polarizations cause greater heating than LP. This effect can be rationalized in terms of the direction between the optically induced magnetization, $M_{IFE}$, and the applied magnetic field, $B_{app}$. When $M_{IFE}$ and $B_{app}$ are anti-parallel, as for RHCP, the optically induced circular electron motion is opposite the direction favored by the Lorentz forces from the external magnetic field, resulting in an increase of damping and lower optical field enhancement. In support of this picture, Gu.

*et al.* theoretically analyzed the behavior of a free electron gas in a nanoparticle under CP excitation and predicted that the optically induced magnetic moment is enhanced (suppressed) when an external magnetic field is aligned (anti-aligned), due to the Lorentz forces on individual electrons that perturb their circular movement[59]. Theoretical studies of magnetoplasmons also predict decreases in damping when rotating surface charge density waves provide magnetization parallel with externally applied magnetic fields[43].

The difference in heating efficiency during LHCP and RHCP excitation allows an estimate of the strength of the optically induced magnetization, $M_{IFE}$, at hot spots in terms of the magnetization, $M_{ind}$, that results from $B_{app}$ (see full calculations in Supporting Information). Assuming that the temperature increase compared to LP excitation is linearly proportional to the net magnetization $M_{ind} + M_{IFE}$ we determine an "effective" magnetic field, $B_{eff}$, at hot spots to be 1.3T for the highest incident power of $1.45 \times 10^9$ W/m$^2$ and 0.67 ellipticity. Note that $B_{eff}$ is not the magnetic field produced by optically exciting the nanostructure, but rather, corresponds to the field strength of a hypothetical external magnet that would produce the same magnetization in the dark as observed during CP optical excitation with no $B_{app}$. This estimate also assumes that $M_{ind}$ and $M_{IFE}$ are either aligned or anti-aligned, though their orientation may be more complex microscopically[40]. When normalized for optical power density, the observed magnetization is in good agreement with our previous time-resolved studies of ensembles of Au colloids[42] (Table S1).

Finally, we comment that the plasmon damping can also be estimated directly by fitting to the Stokes side eR spectrum (green box, Fig. 2b). As discussed in detail in a recent report from our laboratory[53], the eR spectrum reflects the approximately Lorentzian distribution of non-thermal electron-hole pairs that have been generated during the plasmon damping process, i.e., the natural linewidth of the excited plasmon. The fitted damping observed under LP ($1.42 \times 10^9$ W/m$^2$) was 34.1 meV and the lowest damping observed under CP ($1.42 \times 10^9$ W/m$^2$, ellipticity of 0.94) was 31.6 meV. These values can equivalently be reported as a plasmon dephasing time of 19.3 fs (LP) or 20.8 fs (CP) and are comparable to values commonly reported in ultrafast transient absorption studies of Au nanostructures[21]. While this fitted estimate of the ellipticity-dependent change in damping is somewhat smaller compared to the estimate based on computational modeling of the sample backscattering study discussed above, both measures consistently indicate a significant decrease of plasmon damping during CP excitation.

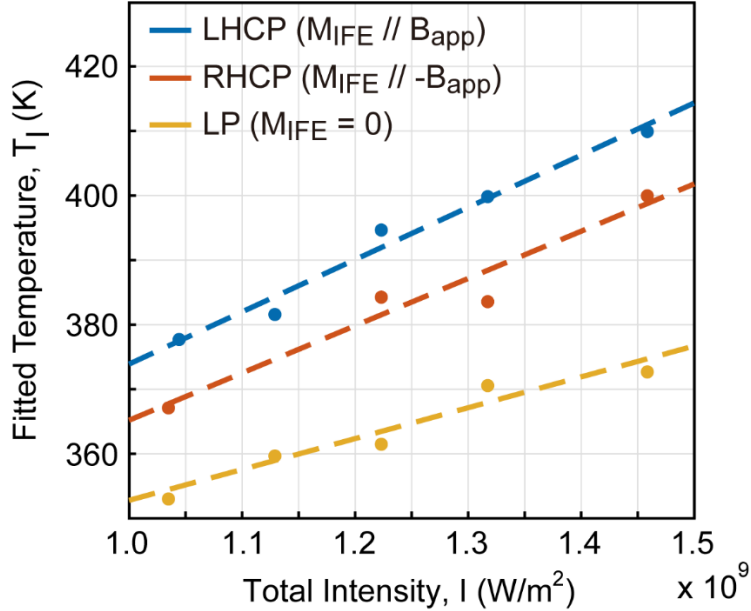

**Fig. 6. Photothermal heating under an external magnetic field and different helicities.** Fitted nanostructure temperature ($T_l$) as a function of total intensity ($I$) with $B_{app}$ = 0.2 T and variable excitation helicity. The magnitude of ellipticity is 0.67 for both LHCP and RHCP. The linear fits to each temperature trend are LHCP: $T_l\,(K) = 8.1 \times 10^{-8} I + 293$; RHCP: $T_l\,(K) = 7.3 \times 10^{-8} I + 292$; LP: $T_l\,(K) = 4.8 \times 10^{-8} I + 305$.

**Conclusion**

We have demonstrated the ability to modulate plasmon damping in achiral plasmonic gold nanodisk arrays by controlling incident light ellipticity. Confocal mapping revealed that CP excitation leads to enhanced efficiency for backscattering, consistent with an overall decrease of damping. A dual-beam Raman thermometry technique quantified localized heating in samples. We observe more efficient photothermal heating when the ellipticity of incident light increases, regardless of helicity (RHCP or LHCP), indicating greater field enhancement at hot spots. The simultaneous increase of scattering *and* absorption is a telltale signature of decreased damping in plasmonic absorbers. In comparison, under an external magnetic field, RHCP and LHCP excitation provide different amounts of heating. This behavior suggests that the microscopic origin of decreased damping is the interaction between the optically driven coherent electron motion and Lorentz forces from DC magnetic fields, whether magnetic fields are optically induced or externally applied. Our results provide further insight into electron dynamics inside plasmonic nanostructures during CP excitation and suggest multiple new strategies for controllably modulating heating, magnetization, reflectance, damping, and related photophysical effects.

## Methods

### Nanostructure Fabrication

Prior to fabrication, a silicon substrate was cleaned using a combination of base piranha and UV-ozone. A 100 nm gold mirror was then thermally evaporated (Lesker PVD electron-beam evaporator) onto the silicon substrate. A 38 nm thick $Al_2O_3$ dielectric layer was then deposited on the gold mirror by RF sputtering (Lesker PVD RF sputterer) and the thickness was determined using ellipsometry. Next, 950 PMMA A4 was spin-coated onto the $Al_2O_3$ as the electron beam resist layer. Electron-beam lithography (TESCAN MIRA3 EBL) was performed to pattern the 100 µm × 100 µm nanodisk array into the e-beam resist. After development, a 5 nm chrome adhesion layer was thermally deposited on the surface of the exposed PMMA, followed by a 100 nm layer of gold. Finally, liftoff was performed in acetone using a combination of pipet pumping and sonication, leaving only the nanodisk array on the surface of the substrate. The morphology of gold nanodisks array is confirmed by SEM (Fig. 1f).

### Raman Spectroscopy

Raman spectra were taken using a confocal microscope system (Witec RA300) and spectrometer (UHTS300, grating = 300 g/mm). For Raman spectral mapping, the excitation source was 532 nm CW Nd:YAG laser and the Raman spectra were collected by a 100x objective (Zeiss EC Epiplan Neofluar, NA = 0.9, WD = 0.31mm). A pair consisting of a holographic 532 nm notch filter (RayShield Coupler, Witec) and a 532 nm notch filter (NF533-17, Thorlab) were added to prevent saturation of the spectrometer. The obtained Raman spectra were corrected by the transmission spectra of the notch filter. The resolution of the Raman map was 50 nm in lateral dimensions (both x and y direction) and 100 nm in the z direction. The optimal z height was determined by maximizing the Raman Stokes signal. The 532 nm backscattering efficiency under different ellipticity was obtained by referencing to a silver mirror.

For the dual-beam Raman spectroscopy experiment, two 532 nm CW laser were used. Beam 2 was coupled through free space, and was a 532 CW diode laser with a spot size of 2 $um^2$ on the sample. The ellipticity of Beam 2 was controlled by a half waveplate and a quarter waveplate optically in series. The function of Beam 2 was to generate circular currents in the gold nanostructures and actively tune the damping constant. The difference in the highest achievable ellipticity for LHCP and RHCP was a result of the limitation of the beam splitter. Beam 1 was coupled through a fiber coupler (Rayshield coupler) and was sourced by a 532 nm CW Nd:YAG laser, with a spot size of 0.55 $µm^2$ when focused. Beam 1 always had lower intensity than Beam 2 and was linearly polarized. Both beams were focused by a 100x objective (Zeiss EC Epiplan Neofluar, NA = 0.9, WD = 0.31mm) on the gold nanostructures. Below the sample, there was a slot for inserting a magnet, which has magnetic field parallel to the incident light (pointing downward) with magnetic field strength = 0.2 T near the surface of gold nanostructure.

**Acknowledgements**
We thank Prof. Luat Vuong for helpful discussions. This work was funded in part by the National Science Foundation (Grant DMR-2004810). M.S. also acknowledges support from the Welch Foundation (A-1886).

**Author contributions**
O.H.-C.C. and B.Z. carried out the measurements and analyzed the data. O.H.-C.C. drafted the manuscript. B.Z. and performed the simulations and drafted the SI. Z.B. fabricated the nanostructures. D.H.S. and M.S. supervised the project and participated in the analysis of the data.

**Supporting information for**
Active tuning of plasmon damping via light induced magnetism

**Author list:**
Oscar Hsu-Cheng Cheng[1]*, Boqin Zhao[1]*, Zachary Brawley[3], Dong Hee Son[1,2], Matthew Sheldon[1,3]

[1]Department of Chemistry, Texas A&M University, College Station, TX, USA.
[2] Center for Nanomedicine, Institute for Basic Science and Graduate Program of Nano Biomedical Engineering, Advanced Science Institute, Yonsei University, Seoul, Republic of Korea
[3]Department of Material Science and Engineering, Texas A&M University, College Station, TX, USA.
*These authors contributed equally to this work.
e-mail: dhson@chem.tamu.edu; sheldonm@tamu.edu

## 1. Full wave electromagnetic simulations

### 1.1 Simulations Setup

Three-dimensional full wave electromagnetic simulations were performed using finite difference time domain (FDTD) methods (Lumerical, Ansys Inc.). A 100 nm height and 218 nm radius (determined by the SEM image) Au nanodisk on a 40 nm thick $Al_2O_3$ layer (with a 5 nm Cr adhesion layer in between) on top of Au film defined the simulation geometry. Periodic boundary conditions were applied on all sides perpendicular to the substrate with a simulation region of 700 nm x 700 nm, representing the periodicity of the nanostructure. A plane wave source was injected from above, normal to the substrate. The backscattered radiation was collected by a 2-D monitor placed above the plane wave source.

The refractive index of $Al_2O_3$ was calculated as a linear combination of the experimental index values from Palik[1] for $Al_2O_3$ and Al, so that the imaginary part at 532 nm is close to 0.2, to account for excess Al in the layer during nanostructure fabrication as determined in control studies. The permittivity values for Au was modeled with the analytical function based on Drude-Lorentz model[2]:

$$\varepsilon(\omega) = 1 - \frac{f_0 \omega_{p,0}^2}{\omega^2 + i\Gamma_0 \omega} + \sum_{j=1}^{j_{max}} \frac{f_j \omega_{p,j}^2}{\omega_j^2 - \omega^2 - i\Gamma_j \omega} \quad (S1)$$

where $\varepsilon(\omega)$ is the relative permittivity, $\Gamma_0$ is the intraband damping constant, $\Omega_p = \sqrt{f_0} \omega_{p,0}$ is the plasma frequency associated with intraband transition. $j_{max}$ is the number of Lorentz oscillators (in our case, 5) with strength $f_j$, frequency $\omega_j$ and damping constant $\Gamma_j$ for each individual oscillator.

In order to systematically study the effect of modulation of damping on optical properties observed in the experiment, we studies the dependence on the intraband damping term $\Gamma_0$, and all the interband damping terms $\Gamma_j$ in equation (S1). Bulk Au values were used for simulations representing LP excitation in the experiment. These damping terms were reduced up to 50% of their original values to generate new permittivity values to represent the CP excitation in the experiment, applied to the Au film and the entire Au nanodisk.

### 1.2 Backscattering simulations

The confocal backscattering map in the experiment (Fig. 3) shows the spatial dependence of backscattering on the nanodisk array. In order to reproduce this spatial dependence behavior in the simulation, we mapped out the distribution of Poynting vector along z direction (Pz) of the scattered field on a x-y plane slightly above the top surface of the nanodisk. We then averaged around each point with a Gaussian-like decay which approximates the effect of Gaussian beam excitation in the experiment (Fig. S1a). The backscattering intensity is the largest near the center of the nanodisk and lowest between the gaps. When comparing the backscattering spatial distribution at different damping (Fig. S1b), it is clear that decreasing damping increases the backscattering intensity along the black dash line in Fig. S1a, which is consistent with experimental results (Fig. 3c, f).

However, with 50% damping modulation, the magnitude of backscattering change predicted by the simulation was not as large as that observed in the experiment at all locations. We hypothesize that the damping modulation is highly spatially non-uniform in the experimental system with maxima near hot spots, unlike the assumption of uniform damping modulation across the nanostructure in the modeling. A spatial dependent change in damping in experiments is likely a key factor that is not accounted for in these simulations, limiting the quantitative accuracy of the result. Nonetheless, the modeling corroborates that the observed experimental trend is consistent with decreased damping in the metal.

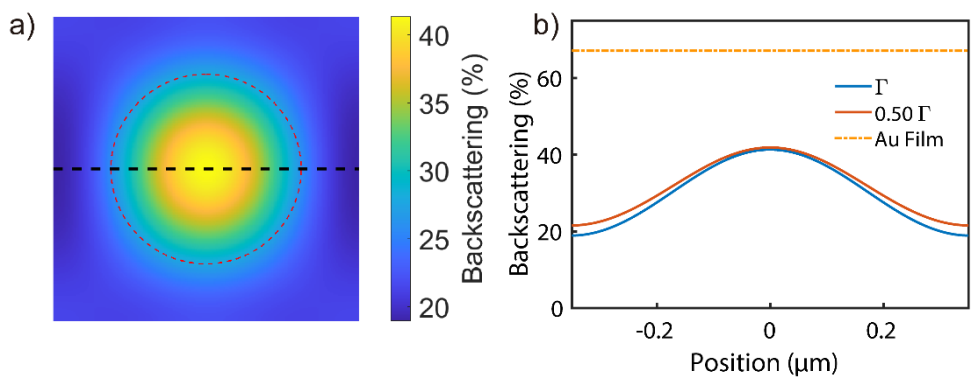

**Fig. S1** a) Simulated backscattering map on the x-y plane above the nanodisk top surface with Gaussian-like weighting. The red line circles region of the nanodisk. b) Backscattering intensity plotted along the black dashed line in a), at bulk damping and 50% damping, as well as the backscattering from an Au film for reference.

In addition to the backscattering map, the overall backscattering intensity from the nanostructure, essentially the average of the spatial dependent backscattering intensity, was also simulated (Fig. S2). The overall value also increases by around 0.5% with damping values decreased by half, further highlighting the increased backscattering efficiency with decreasing damping.

If we label the overall backscattered percentage of power as R, the absorbed percentage of power would be 1-R, due to the opacity of the nanostructure. Therefore, the absorption spectrum of the nanostructure can be computed accordingly. Fig. S3 shows the computed absorption spectrum of the nanostructure. Although it does not exactly match the experimental absorption spectrum in Fig. 2a, but the general shape is very similar.

We emphasize that simulation does not model any nonlinear effects. If we only change the ellipticity of the excitation source in the simulations, without modulating Au permittivity values, the absorption and the backscattering intensity values of the nanostructure remain constant. This suggests that the difference in absorption and scattering seen in the simulation must come from some nonlinear effects, such as modulation of damping.

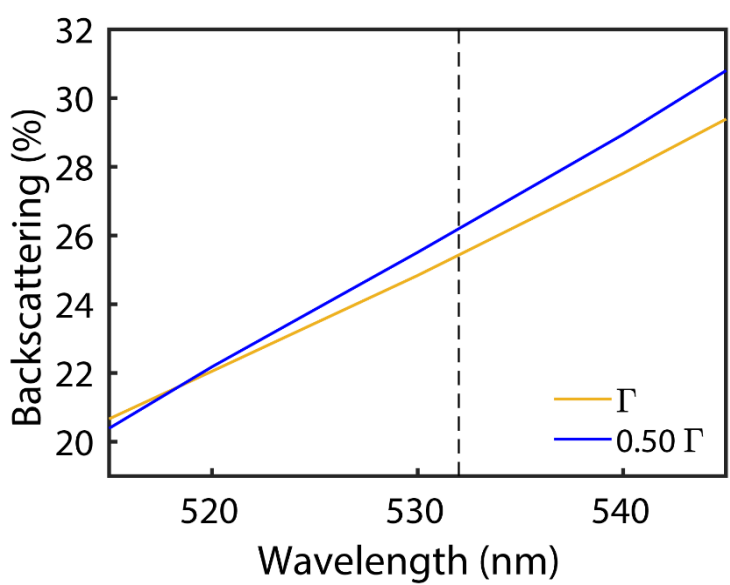

**Fig. S2** Overall backscattering efficiency of gold nanodisk arrays with damping of 100% and 50% of bulk damping.

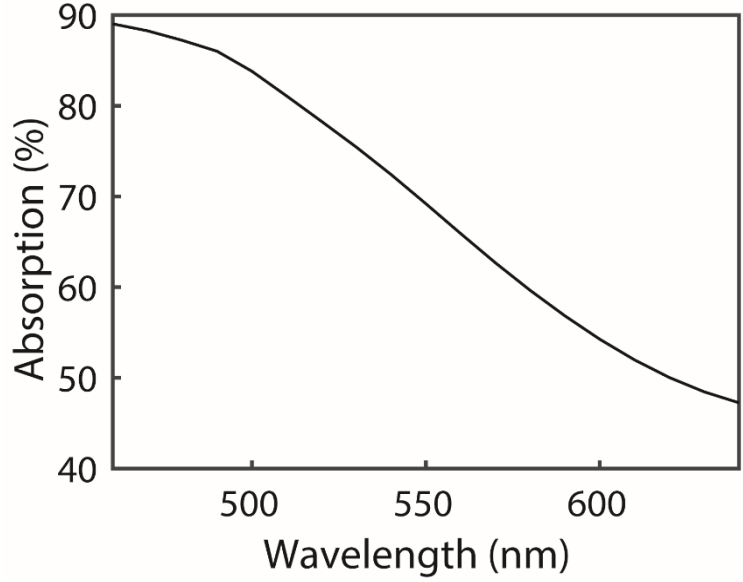

**Fig. S3** Simulated absorption spectrum of gold nanodisk arrays with the bulk damping constant.

1.3 Plasmonic heating simulations

We mentioned in the main article that the local power density for heating, $q(x,y,z)$, depends on field enhancement as:

$$q(x,y,z) = \frac{1}{2}Im[\varepsilon(x,y,z)]\varepsilon_0|E(x,y,z)|^2 \quad (S3)$$

where $E(x,y,z)$ is the local electric field. We simulated the local electric field distribution $E(x,y,z)$ on the nanodisk with damping modulation to verify the experimentally observed increase in photothermal heating with decreasing damping. Here, we mapped the electric field intensity ($|E|^2$) distribution on the top surface of the nanodisk, multiplied by the imaginary permittivity, which is proportional to the local heating power. Fig. S4 shows the calculated results for both bulk damping and 50% damping. At 50% damping, the local heating at hot spots is significantly higher than the bulk damping case.

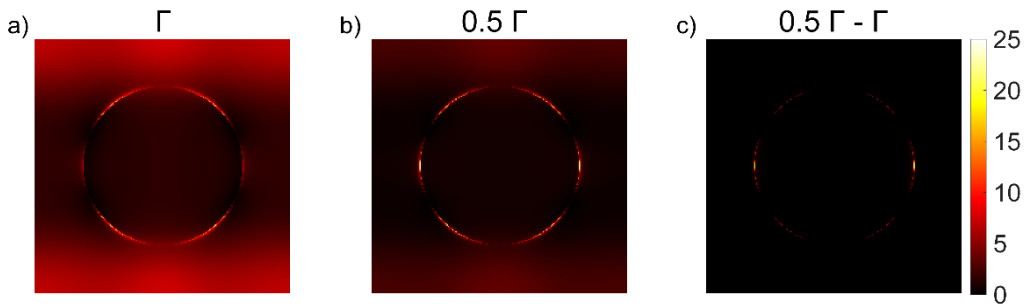

**Fig. S4** Heating map on the top surface of the nanodisk a) with bulk damping, b) with 50% damping and c) the difference between 50% damping and bulk damping.

## 2. Raman fitting methodology

The method used in the manuscript to extract the temperature of the plasmonic nanostructure was adopted from Xie et al[3] and from many of our previous reports[4-7]. Basically, the anti-Stokes electronic Raman scattering signal from the plasmonic nanostructure is associated with the thermalized electron distribution that is approximately thermally equilibrated with the metal lattice. Therefore, the spectral intensity of the anti-stokes spectrum $S(\Delta\omega)$ can be characterized with a Bose-Einstein distribution:

$$S(\Delta\omega) \propto C \times D(\Delta\omega) \times \frac{1}{\exp\left(-\frac{hc\Delta\omega}{k_B T}\right)-1} \quad \text{(S3)}$$

where $h$ is Plank constant, $c$ is the speed of light, $\Delta\omega$ is the wavenumber (negative for anti-Stokes), $k_B$ is Boltzmann constant, and $T$ is the temperature of the sample. $C$ is a scaling factor. $D(\Delta\omega)$ is a correction factor proportional to the photonic density of states (PDOS) of the nanostructure. In practice, the PDOS is approximated with reflection[4-6], absorption, or dark-field scattering[8] spectrum. However, there is not a direct experimental measurement of the PDOS. Therefore, we follow the procedure of Xie et al[3] and have normalized the Raman spectrum at an unknown temperature $T_l$ to a reference Raman spectrum at a known temperature $T_0$, to remove the dependence of $D(\Delta\omega)$:

$$\frac{S(\Delta\omega)_{T_l}}{S(\Delta\omega)_{T_0}} = \frac{C_1 \times D(\Delta\omega) \times \left(\exp\left(\frac{-hc\Delta\omega}{k_B T_0}\right)-1\right)}{C_2 \times D(\Delta\omega) \times \left(\exp\left(\frac{-hc\Delta\omega}{k_B T_l}\right)-1\right)} = C \times \frac{\exp\left(\frac{-hc\Delta\omega}{k_B T_0}\right)-1}{\exp\left(\frac{-hc\Delta\omega}{k_B T_l}\right)-1} \quad \text{(S4)}$$

The reference spectrum is usually chosen to be the spectrum at lowest illumination intensity, assuming no laser heating, and therefore the sample temperature is assumed to be at room temperature.

By correctly normalizing the spectral counts to the illumination power and integration time, $C$ is theoretically a value of 1. Therefore, the only parameter to solve for is the unknown temperature $T_l$. In practice, fitting accuracy is improved if $C$ is a free fit parameter (usually very close to 1) to account for small changes in the collection efficiency during the course of the measurement.

### 3. The accuracy of Raman-fitted temperatures

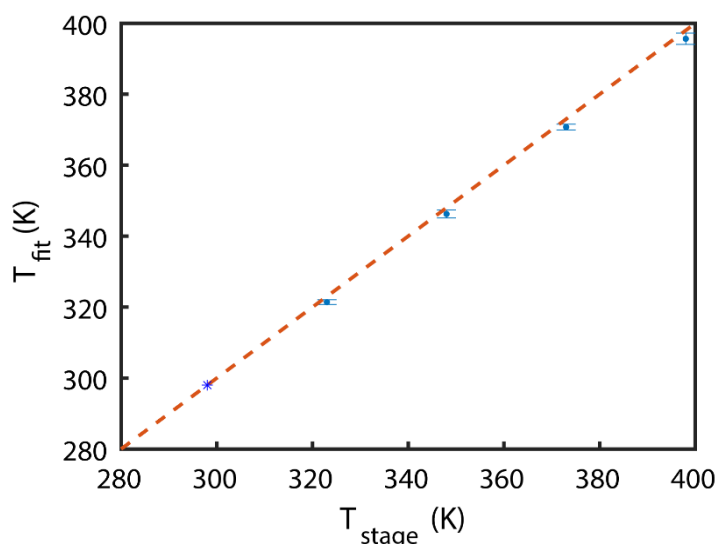

**Fig. S5 Fitted temperature of the nanostructure compared to the thermal stage temperature.** Blue asterisk with error bar: average fitted temperature and standard deviation. Red dotted line: reference line with $T_{stage}$ = $T_{fit}$.

We have shown that the sample temperature extracted from the Raman spectrum with equation (3) is physically accurate. The same nanostructure sample used in the manuscript was heated to a pre-set temperature on a thermal stage (TS1500, Linkam). Raman spectra were taken at different thermal stage temperatures with only beam 1 present. The intensity of beam 1 was set to be very low for negligible laser heating. The Raman spectrum collected at 298 K thermal stage temperature was chosen for the reference spectrum and $T_0$ was set to be 298 K in the fitting.

The fitted temperatures were compared to the corresponding thermal stage temperatures, as shown in Fig. S5. Standard deviation (the blue bar around the data points) was calculated based on the fitted temperatures for several spectra at the same thermal stage temperature, which is around 1 K, showing the high precision of the measurements. The data points lie close to the red reference line with $T_{stage}$ = $T_{fit}$, confirming our ability to accurately report the nanostructure temperatures with anti-Stokes Raman thermometry.

### 4. Determination of optically-induced magnetism $M_{IFE}$ at hot spots

Although bulk Au has been reported to display diamagnetism, there have been numerous reports pointing to observed paramagnetic, or even ferromagnetic behavior in various nanoscale Au systems[9-12]. According to the theory of the IFE[13, 14] and by studying the direction of magnetization in this report, we concluded that the Au nanodisk arrays behave paramagnetically (under circularly polarized optical pumping). However, the (volume) susceptibility is not known. Here, we may label it symbolically with $\chi_V$ and carry it through our subsequent calculations. In addition, our previous experimental study[15] also indicates a paramagnetic behavior in nanoscale Au. Therefore, we model the magnetization of Au under an external magnetic field (in the dark) as follows:

$$B_{\text{app}} = \mu_0 \left(\frac{1}{\chi_V}\right) M_{\text{ind}} \quad (S5)$$

where $\mu_0$ is the vacuum permeability, $B_{\text{app}}$ is the external applied magnetic field and $M_{\text{ind}}$ is the induced magnetization.

The induced magnetization in response to the external magnetic field $M_{\text{ind}}$ is a separate contribution to the sample magnetism in addition to the light-induced magnetism $M_{\text{IFE}}$ (Fig. 1a), and both effects are expected to contribute to the damping experienced by coherently driven, circulating electrons during optical pumping. Therefore, the temperature increase measured in the experiment (Fig. 6, Fig. S6) is assumed to scale linearly with the total induced magnetism $M_{\text{ind}} + M_{\text{IFE}}$.

First, we calculate the strength of $M_{\text{ind}}$ under a 0.2 T applied external magnetic field. By applying equation (S5), with $\mu_0 = 1.257 \times 10^{-6}$ H/m, we can derive that $M_{\text{ind}} = 1.6 \times 10^5 \chi_V$ A/m. We determine $M_{\text{IFE}}$ in the experiment based the fitted temperature data at the highest incident intensity in Fig. 6 (also Fig. S6). Note that the temperature without an external magnetic field, when $M_{\text{ind}} = 0$, would have been 405 K for both LHCP and RHCP. This means the samples experience a temperature increase of ~33 K during CP compared to LP when only the $M_{\text{IFE}}$ alters the damping in the sample. The additional 5 K increase or decrease of temperature (see Fig. S6) during LHCP or RHCP, respectively, is due to the interaction with $M_{\text{ind}}$. Therefore, $M_{\text{IFE}}$ is calculated to be approximately 6.6x larger than $M_{\text{ind}}$, or $1.05 \times 10^6 \chi_V$ A/m, which corresponds to induced magnetic flux density of $1.3 \chi_V$ T. According to equation (S5), this is equivalent to $M_{\text{ind}}$ that would be produced under a 1.3 T external applied magnetic field, which we label as $B_{\text{eff}}$.

Lastly, we can calculate the magnetic moment per Au atom $m_{\text{Au}}$ based on the relationship

$$M_{\text{IFE}} = \frac{N_{\text{Au}}}{V} m_{\text{Au}} \quad (S6)$$

In an Au crystal lattice, the number density of Au atoms $\frac{N_{\text{Au}}}{V}$ is 58.9 nm$^{-3}$. Therefore, $m_{\text{Au}} = 1.9 \chi_V \mu_B$.

We compare our results with our previous experimental study on Au nanocolloids, as well as two experiment measurements on Au film, summarized in Table. S1, assuming that the IFE magnetism is linearly proportional to the incident optical power (as suggested by theoretical studies on the IFE[13, 14, 16]). Our analysis in this study only reveals the magnetic behavior at hot spots, because most of the Raman signal comes from hot spots on the nanodisk. In Ref. [15], the magnetic moment per Au atom was averaged over the entire nanoparticle, which could be the reason for a slightly smaller value for induced magnetic moment per atom compared to this study. Both nanoscale Au system possess excitation intensity normalized magnetic moment per atom ($m_{\text{Au}}/I$) of almost 4 orders of magnitude larger than Au film, indicating an enhancement of the IFE phenomenon in nanoscale compared to bulk.

For reference, if $\chi_V$ is on the order of $10^{-5}$ (a typical value for bulk Au), the IFE induced magnetic field at hot spots during in the study is estimated to be on the order of $10^{-5}$ T.

**Table. S1 Comparison between induced magnetic moment per atom due to the IFE in different reports.**

|  | System | Excitation Intensity $I$ (W/m²) | $M_{\text{IFE}}$ (A/m) | $m_{\text{Au}}$ ($\mu_B$) | $m_{\text{Au}}/I$ ($10^{-10}\mu_B/(\frac{W}{m^2})$) |
|---|---|---|---|---|---|
| This work | Au nanodisk array | ~$10^9$ (CW) | $1.05 \times 10^6 \chi_V$ | $1.9\chi_V$ | [b]$19\chi_V$ |
| Ref. [15][15] | Au nanoparticle colloid | ~9×10¹³ (pulsed) | $3.0 \times 10^4 \chi_V$ | $2.8 \times 10^4 \chi_V$ | $3.1\chi_V$ |
| Ref. [17][17] | Au film | ~13×10¹³ (pulsed) | [a]$4.5 \times 10^6 \chi_V$ | $8.2\chi_V$ | $6.3 \times 10^{-4}\chi_V$ |
| Ref. [18][18] | Au film | ~13×10¹³ (pulsed) | [a]$1.1 \times 10^7 \chi_V$ | $20\chi_V$ | $1.5 \times 10^{-3}\chi_V$ |

[a] Several assumptions were made to calculate this value. The path length in Au film is estimated by the skin depth (1/absorption coefficient). The verdet constant of Au film is extracted from Ref. [17][19]. The induced magnetization was also normalized to the excitation frequency. $\chi_V$ of Au film was used as bulk Au value.

[b] At hot spots.

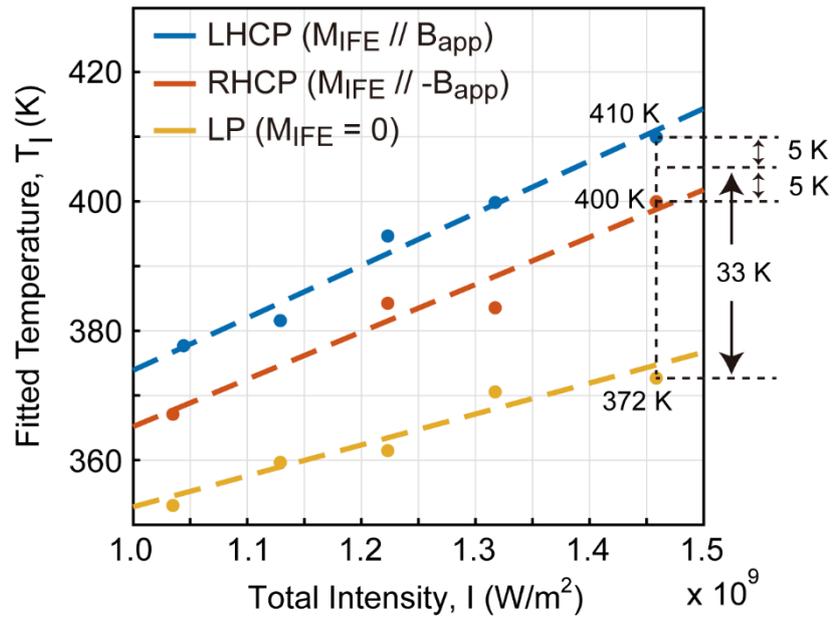

**Fig. S6** Diagram to aid induced magnetic field calculation.

## 5. Optical setup

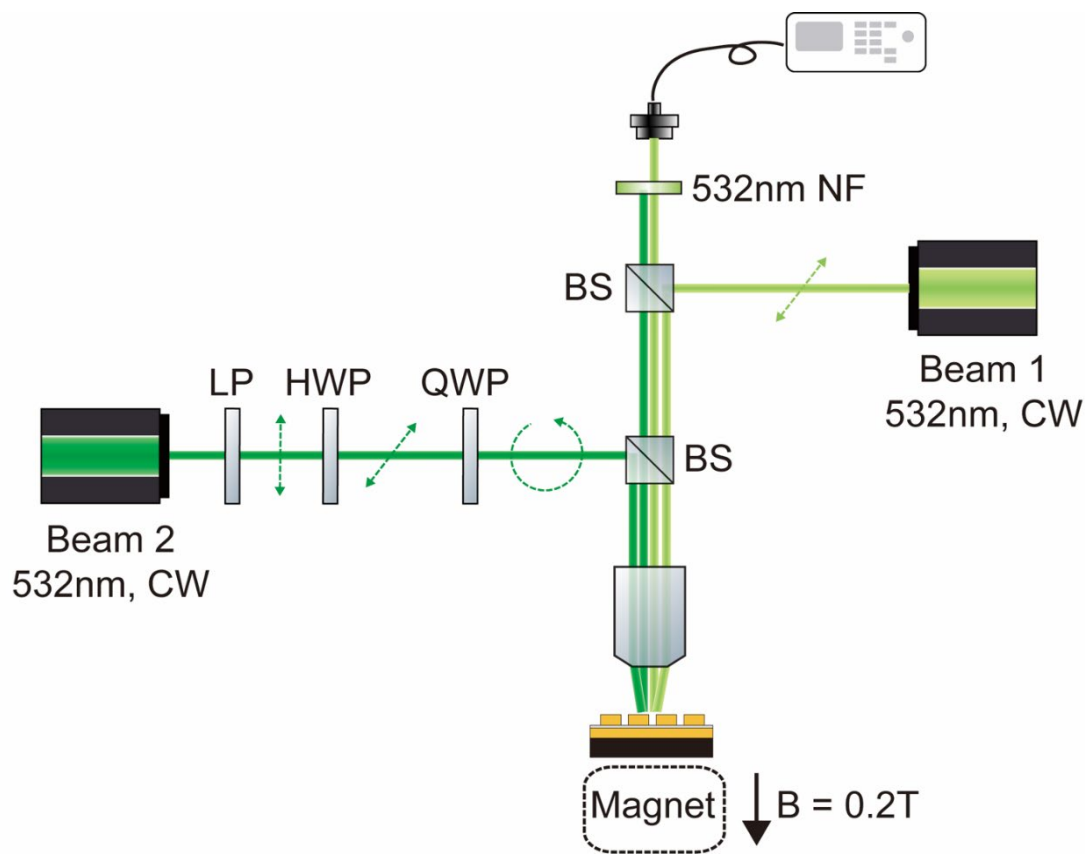

**Fig. S7. Optical setup for dual-beam Raman spectroscopy.** CW: continuous wave; LP: linear polarizer; HWP: half waveplate; QWP: quarter waveplate; BS: beam splitter.

## 6. Definition of ellipticity

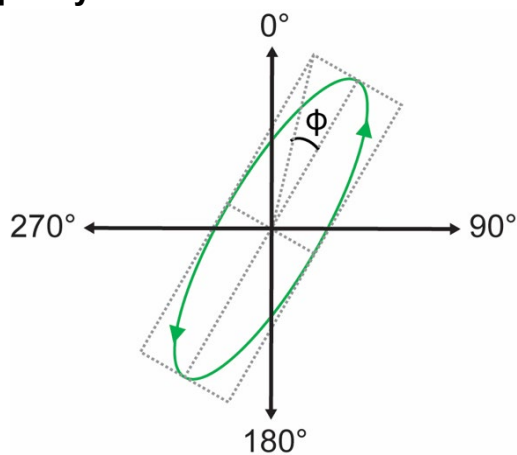

**Fig. S8. Ellipticity explanation.** This figure shows a left-handed elliptically polarized light. φ is the ellipticity angle. Ellipticity is defined as the intensity ratio between the short and long axis. Ellipticity angle = 0, -45º, +45º or Ellipticity = 0, 1, -1 represent LP, LHCP, RHCP, respectively.